\documentclass[pra,reprint, superscriptaddress,floatfix]{revtex4-1}

\usepackage[utf8]{inputenc}
\usepackage{amsmath}
\usepackage{graphicx}
\usepackage{dcolumn}
\usepackage{wrapfig}
\usepackage{amsmath}
\usepackage{amssymb}
\usepackage{setspace}
\usepackage{float}
\usepackage{comment}
\usepackage{xcolor}
\usepackage{textgreek}
\usepackage{hyperref}
\usepackage{tabularx}
\usepackage{array}
\usepackage{longtable}

\newcommand{\rqul}{RF Quantum Upconverter}
\newcommand{\rqu}{RQU}

\date{July 2024}

\begin{document}

\title{Quantum metrology of low frequency electromagnetic modes with frequency upconverters}

\author{Stephen E. Kuenstner}
\affiliation{Department of Physics, Stanford University, 382 Via Pueblo, Stanford, CA 94305.}

\author{Elizabeth C. van Assendelft}
\affiliation{Department of Physics, Stanford University, 382 Via Pueblo, Stanford, CA 94305.}

\author{Saptarshi Chaudhuri}
\affiliation{Department of Physics, Princeton University, Jadwin Hall Washington Road, Princeton, NJ 08544.}

\author{Hsiao-Mei Cho}
\affiliation{SLAC National Accelerator Laboratory, 2575 Sand Hill Rd, Menlo Park, CA 94025}

\author{Jason Corbin}
\affiliation{Department of Physics, Stanford University, 382 Via Pueblo, Stanford, CA 94305.}

\author{Shawn W. Henderson}
\affiliation{SLAC National Accelerator Laboratory, 2575 Sand Hill Rd, Menlo Park, CA 94025}

\author{Fedja Kadribasic}
\affiliation{Department of Physics, Stanford University, 382 Via Pueblo, Stanford, CA 94305.}

\author{Dale Li}
\affiliation{SLAC National Accelerator Laboratory, 2575 Sand Hill Rd, Menlo Park, CA 94025}

\author{Arran Phipps}
\affiliation{Department of Physics, 
California State University East Bay, 25800 Carlos Bee Blvd, North Science 231, Hayward, CA 94542}

\author{Nicholas M. Rapidis}
\affiliation{Department of Physics, Stanford University, 382 Via Pueblo, Stanford, CA 94305.}

\author{Maria Simanovskaia }
\affiliation{Department of Physics, Stanford University, 382 Via Pueblo, Stanford, CA 94305.}

\author{Jyotirmai Singh}
\affiliation{Department of Physics, Stanford University, 382 Via Pueblo, Stanford, CA 94305.}

\author{Cyndia Yu}
\affiliation{Department of Physics, Stanford University, 382 Via Pueblo, Stanford, CA 94305.}

\author{Kent D. Irwin}
\affiliation{Department of Physics, Stanford University, 382 Via Pueblo, Stanford, CA 94305.}
\affiliation{SLAC National Accelerator Laboratory, 2575 Sand Hill Rd, Menlo Park, CA 94025}

\begin{abstract}
We present the \rqul{} (\rqu{}) and describe its application to quantum metrology of electromagnetic modes between dc and the Very High Frequency band (VHF) ($\lesssim$300MHz). The \rqu{} uses a Josephson interferometer made up of superconducting loops and Josephson junctions to implement a parametric interaction between a low-frequency electromagnetic mode (between dc and VHF) and a mode in the microwave C Band ($\sim$ 5GHz), analogous to the radiation pressure interaction between electromagnetic and mechanical modes in cavity optomechanics. We analyze \rqu{} performance with quantum amplifier theory, and show that the \rqu{} can operate as a quantum-limited op-amp in this frequency range. It can also use non-classical measurement protocols equivalent to those used in cavity optomechanics, including back-action evading (BAE) measurements, sideband cooling, and two-mode squeezing. These protocols enable experiments using dc--VHF electromagnetic modes as quantum sensors with sensitivity better than the Standard Quantum Limit (SQL). We demonstrate signal upconversion from low frequencies to microwave C band using an \rqu{} and show a phase-sensitive gain (extinction ratio) of $46.9$\;dB, which is a necessary step towards the realization of full BAE.
\end{abstract}

\maketitle

\section{Introduction}
\label{sec:Introduction}

The field of Circuit Quantum Electrodynamics (Circuit QED) has made impressive strides in harnessing the quantum-mechanical properties of superconducting circuits operating in the microwave frequency regime (typically several GHz) \cite{Blais2007}. The techniques of Circuit QED 
have advanced to the point that detecting \cite{Johnson2010, chen2011} and coherently manipulating \cite{nakamura1999} a single microwave quantum are routine operations, and individual control over arrays of dozens of interacting quantum circuits is possible \cite{arute2019}. Much of this progress has been driven by the desire to build a universal quantum computer capable of performing calculations that would be impractical on any classical computer.

The techniques of Circuit QED do not extend directly to lower frequencies, however. Recently, there has been growing interest in adapting quantum metrology techniques to lower frequency electromagnetic modes, typically at frequencies between dc and the Very High Frequency (VHF) band below 300 MHz. Quantum metrology of low-frequency modes could offer a sensitivity advantage over classical sensors, enabling experiments including dark-matter searches \cite{brouwer2022proposal}, low-frequency nuclear spin metrology \cite{clarkeSpinNoise}, some astronomical measurements \cite{zmuidzinas2003thermal}, and low-frequency magnetometry, outperforming a dc SQUID in certain applications. 

One approach for quantum RF metrology that has been developed at these frequencies is to use a qubit to cool the MHz resonator to its ground state and stabilize a Fock state with a small number of photons \cite{gely2019observation}. While this approach can be used in principle to measure individual signal photons entering the resonator, it does not provide a way to discriminate incoming signal photons from background thermal photons, which limits its usefulness for certain measurements.

For example, searches for axion or axion-like dark matter at mass below 1\;\textmu eV must detect or rule out yoctowatt-scale or smaller electromagnetic signals over many decades in frequency, spanning from $\sim$100Hz to $\sim$300MHz \cite{Chaudhuri2015, phipps2020, Ouellet2019, crisosto2020, Garcon2018, Nguyen2019, brouwer2022proposal}. These signals can be used to excite an electromagnetic resonator. If the photon number state of the resonator is then measured, the signal-to-noise ratio is limited by the random entry or departure of background thermal photons from the resonator, since photon-counting techniques lack the frequency resolution to distinguish thermal and signal photons. An alternative approach that applies quantum techniques to measure the dark-matter-induced voltage, rather than the photon number, allows the signal frequency to be determined. Frequency information is important as these circuits carry useful information significantly detuned from their resonant frequency. Far from the resonant frequency, thermal fluctuations are suppressed to below the level of a single photon per second per Hz of bandwidth \cite{chaudhuri2018, chaudhuri2019}. This off-resonant signal information can be accessed using continuous-variables readout techniques operating beyond the Standard Quantum Limit (SQL). In this case, improving the readout performance does not substantially improve the signal to noise ratio (SNR) on resonance (which is limited by thermal fluctuations), but it allows constant SNR to be maintained over a much broader bandwidth, dramatically increasing the axion search rate.

In this work, we describe the \rqul{} (\rqu{}), a device that mimics the radiation-pressure interaction in cavity optomechanics, but replaces the low-frequency mechanical mode with a low-frequency electromagnetic mode. The \rqu{} uses the nonlinearity of Josephson junctions to upconvert signals from the sensor frequency to microwave frequencies using a three-wave mixing interaction. We experimentally demonstrate three-wave mixing with an \rqu{} in $\S$\ref{sec:upconversion-demo}.

This upconversion paradigm allows the \rqu{} to take advantage of several mature microwave Circuit QED technologies, including high coherence microwave resonators \cite{Reagor2016}, Josephson Parametric Amplifiers (JPAs) \cite{Roy2016}, and microwave squeezers \cite{CastellanosBeltran2008}, while extending the frequency range of quantum measurement techniques to lower frequencies. They also enable the use of quantum-limit evading techniques equivalent to those used in cavity optomechanics, including back-action evading (BAE) measurements\cite{Clerk2008}. After analyzing BAE using the \rqu{}, in $\S$\ref{sec:upconversion-demo} we experimentally demonstrate a phase-sensitive gain of $46.9$\;dB, which is a significant step towards the realization of full BAE.

\section{Analogy between cavity optomechanics and op-amp mode amplification}
\label{sec:generic-RQU}

\subsection{Upconverter Hamiltonian}
\label{subsec:upconverter-Hamiltonian}

To evaluate the \rqu{} as a tool for quantum metrology, we use a model in which both the \rqu{} and its input circuit are quantized, with an interaction Hamiltonian that couples the modes. This Hamiltonian is exactly analogous to that of cavity optomechanics, but the mechanical mode is replaced with an electromagnetic mode, referred to in this section as the ``low-frequency mode'' to distinguish it from the microwave mode.

Cavity optomechanics treats two bosonic modes at different frequencies: an electromagnetic mode at $\omega_a$ and a mechanical mode at $\omega_b$, with  $\omega_a \gg \omega_b$ \cite{Aspelmeyer2014}. The position operator of the mechanical mode represents the position of a movable mirror that forms one end of the optical cavity. The modes are quantized with ladder operators $\hat{a}, \hat{a}^\dagger$   and $\hat{b}$, $\hat{b}^\dagger$, respectively. The uncoupled Hamiltonian is:
\begin{equation}
\label{eq:h0}
\hat{H}_{0} = \hbar \omega_{a} (\hat{a}^\dagger \hat{a} +1/2) + \hbar \omega_b (\hat{b}^\dagger \hat{b} +1/2),
\end{equation}
In terms of ladder operators, the mirror position is given by:
\begin{equation}
    \label{eq:postion-operator}
    \hat{x} = x_{\rm ZPF} (\hat{b} + \hat{b}^\dagger), 
\end{equation}
where $x_{\rm ZPF}$ is the magnitude of the zero-point position fluctuations. The frequency of photons occupying the optical mode depends on the position of the movable mirror, leading to the parametric optomechanical interaction $\hat{H}_{\rm{int}}^{\rm{OM}}$ between the two modes:
\begin{equation}
\label{eq:hOM}
\hat{H}_{\rm{int}}^{\rm{OM}} = -  \frac{\hbar g_0}{x_{\rm ZPF}} \hat{a}^\dagger \hat{a} \hat{x},
\end{equation}
where $g_0$ is the optomechanical coupling strength, describing the frequency shift of an optical photon due to the position $\hat{x}$ of the mechanical oscillator. 

\begin{figure}
\includegraphics[width=0.95\linewidth]{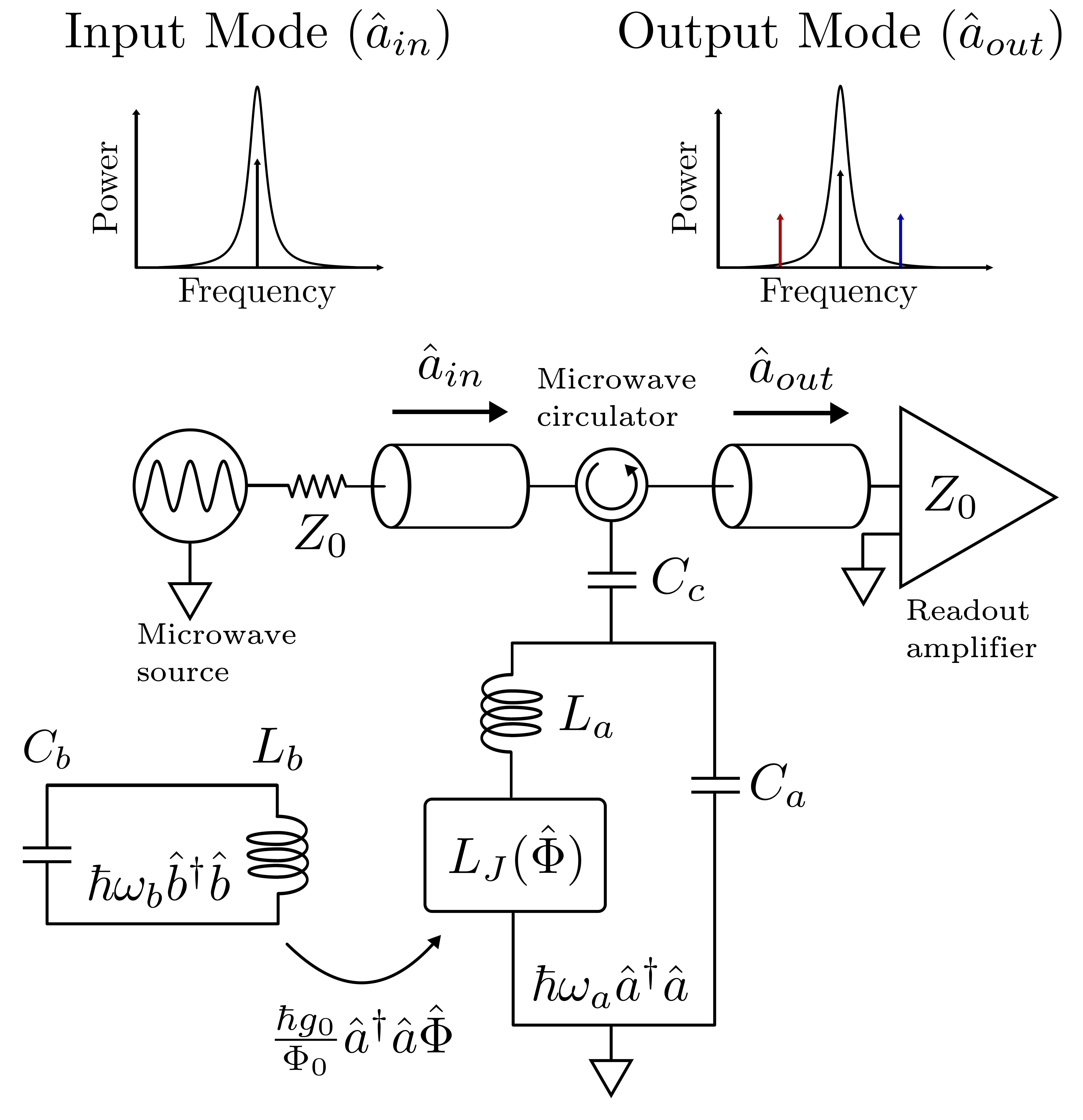}
\caption{A circuit model for an \rqu{}, which inductively couples a dc-VHF signal source (shown here as a resonator formed by $C_b$ and $L_b$) to a tunable Josephson inductance $L_J(\hat{\Phi})$. The tunable inductor is made up of a superconducting interferometer with one or more Josephson junctions (JJs) and one or more loops. The flux $\Phi$ threading the inductor $L_b$ associated with the low-frequency mode also couples through a designable mutual inductance to each of the loops in the JJ interferometer. Thus, $\hat{\Phi}$
changes the inductance $L_J$ presented by the JJ interferometer to the microwave mode, and modifies the resonance frequency of the microwave resonator formed by the interferometer and linear reactances modeled by circuit elements $C_a$ and $L_a$. A coupling capacitance $C_c$, microwave transmission lines and a circulator allow the state of the microwave resonator to be driven and detected via traveling wave modes $\hat{a}_{\rm{in}}$ and $\hat{a}_{\rm{out}},$ respectively. The output mode contains information in sidebands, as shown schematically in the frequency domain.  Low noise amplification by a cryogenic microwave amplifier allows for efficient detection of the output state $\hat{a}_{\rm{out}}$. }
\label{fig:generic-upconverter}
\end{figure}

An optomechanical-style coupling can be realized in microwave superconducting resonant circuits by including a Josephson interferometer whose inductance $L_J(\hat{\Phi})$ depends on the flux $\hat{\Phi}$ in the low-frequency mode. The flux causes the microwave resonance frequency to vary, just as position shifts of the moving mirror cause the optical frequency to vary in the optomechanical setup. This optomechanical-style coupling has been realized in the microwave SQUID multiplexer \cite{mates2008demonstration} although not optimized to approach quantum limits. The dispersive nanoSQUID magnetometer \cite{Levenson-Falk_2013_Dispersive} uses a similar frequency-tunable microwave resonator, but does not use a resonant low-frequency circuit on its input, removing the effects of quantum backaction and limiting the quantum protocols which could be employed. There are also other devices in which a resonator is tuned with a Josephson junction array, including the Asymmetrically Threaded SQUID (ATS) \cite{lescanne2020exponential}. However, in the ATS, the lower frequency signal and higher frequency signal are both coupled as a current drive to the interferometer, and a flux input is used to bias and pump the interferometer for degenerate four-wave mixing. In the \rqu{}, which uses optimechanical-style coupling, the low-frequency signal is applied as a flux to the Josephson interferometer and three-wave mixing is realized. Figure \ref{fig:generic-upconverter} shows a circuit model of such an upconverter with the associated ladder operators.

The uncoupled Hamiltonian of the \rqu{} is exactly the one in equation \ref{eq:h0}, with the phonon ladder operators $\hat{b}, \hat{b}^\dagger$ replaced by photon operators for the low-frequency mode. The microwave and low-frequency modes are represented by harmonic oscillators with frequencies:
\begin{align}
    \label{eq:omega-a}
    \omega_{a}(\hat{\Phi})  = \big((L_a+L_{J}(\hat{\Phi}))(C_a+C_c)\big)^{-\frac{1}{2}},  \\
    \label{eq:omega-b}
    \omega_b  = (L_b C_b)^{-\frac{1}{2}}.
\end{align}
The low frequency mode is inductively coupled to the Josephson interferometer such that the flux threading the low-frequency resonator also threads the Josephson interferometer. Arrays of multiple Josephson junctions and multiple loops can be used with both gradiometric and non-gradiometric coupling to optimize circuit response for different applications. The tunability of $L_J(\hat{\Phi})$ mediates a parametric interaction, with $\hat{\Phi}$ playing the role of the position operator $\hat{x}$. Analogously to equation \ref{eq:postion-operator}, we have:
\begin{equation}
    \label{eq:flux-operator}
    \hat{\Phi} = \Phi_{\rm{ZPF}} (\hat{b} + \hat{b}^\dagger),
\end{equation}
where $\Phi_{\rm{ZPF}}$ is magnitude of the zero point flux fluctuations:
$\Phi_{\rm{ZPF}} = \sqrt{\hbar \omega_b L_b/2}$.

In order to treat the interaction between the modes, we include the perturbation of $L_J$ due to the sensor flux. We analyze the behavior of the \rqu{} in response to small, time-varying flux signals satisfying $|\langle\hat{\Phi}(t)\rangle| \ll \Phi_0$, where $\Phi_0 = h/2e$ is the magnetic flux quantum that sets the periodicity of the interferometer’s response to external magnetic flux. In this regime, we Taylor expand the microwave frequency to first order in the sensor flux, to calculate the shift of the microwave resonance frequency due to flux in the sensor:
\begin{equation}
    \label{eq:H-RQU}
    \hat{H} = \hbar \left(\omega_{a}(0) + \frac{d \omega_a}{d \Phi} \hat{\Phi}\right) (\hat{a}^\dagger \hat{a} +1/2) + \hbar \omega_b (\hat{b}^\dagger \hat{b} +1/2).
\end{equation}
The frequency shift per unit applied flux describes the strength of the interaction between the modes, with:
\begin{equation}
    \label{eq:interaction-strength}
    \frac{d \omega_a}{d \Phi} = \frac{d \omega_a}{d L_J} \frac{d L_J}{d \Phi}.
\end{equation}
The two derivatives on the RHS of equation \ref{eq:interaction-strength} depend on the particular design of the interferometer and low frequency resonator, which we can calculate for a given interferometer design. We can write the upconverter interaction Hamiltonian in equation \ref{eq:H-RQU} in a form analogous to the radiation pressure interaction in equation \ref{eq:hOM}:
\begin{equation}
    \label{eq:H-RQU-int}
    \Hat{H}_{\mathrm{int}}^{\mathrm{RQU}} = - \frac{\hbar g_0}{\Phi_{\rm{ZPF}}} \hat{a}^\dagger \hat{a} \hat{\Phi} = -\hbar g_0 \hat{a}^\dagger \hat{a} (\hat{b}^\dagger + \hat{b}).
\end{equation}
Without loss of generality, we choose the the sign of increasing $\hat{\Phi}$ to yield the minus sign in equation \ref{eq:H-RQU}. Because it involves products of three ladder operators, this interaction describes three-wave mixing. The strength of the optomechanical-style coupling is given by:

\begin{equation}
g_0 \equiv \frac{d \omega_a}{d \Phi} \Phi_{\rm{ZPF}}. 
\end{equation}

\subsection{Input-Output Model}
\label{subsec:input-output}

In order to operate the \rqu{}, we control and detect the state of the microwave resonator, which allows us to infer the state of the low-frequency resonator. The Hamiltonian in equation \ref{eq:H-RQU} only accounts for the interaction between the two modes, and does not include external couplings or dissipation. In order to detect and control the state of the microwave resonator, we couple the microwave resonator to a waveguide that allows microwave photons to escape the cavity for amplification and demodulation.  Finally, the model must also account for the effects of internal dissipation in both the RF and microwave modes.

The total Hamiltonian, accounting for the external coupling, dissipation, and the microwave drive is given by:
\begin{equation}
        \label{eq:htot}
   \hat{H}_{\rm{tot}} = \hat{H}_0 + \hat{H}_{\rm{int}}^{\rm{RQU}} + \hat{H}_\kappa + \hat{H}_\gamma + \hat{H}_{\rm{drive}},
\end{equation}
where $\hat{H}_0 + \hat{H}_{\rm{int}}^{\rm{RQU}}$ describes the dynamics of the isolated \rqu{} system (microwave resonator plus low-frequency resonator, and their interaction), as described in equations \ref{eq:h0} and \ref{eq:H-RQU}. $\hat{H}_\kappa$ captures the effects of loss in the microwave resonator, which is dominated by loss to the strongly coupled readout port. $\hat{H}_\gamma$ describes loss to internal dissipation in the low-frequency resonator. Finally, $\hat{H}_{\rm{drive}}$ accounts for the energy supplied by the external drive tones which probe the \rqu{} state.

The traveling-wave modes used in this section are shown in figure \ref{fig:generic-upconverter}: the microwave resonator is coupled to an ``input'' mode $\hat{a}_{\rm{in}}$ and an ``output'' mode $\hat{a}_{\rm{out}}$ which are used to drive and detect the state of the microwave resonator. The field circulating within the microwave resonator, $\hat{a}$, is referred to as the intra-cavity field. A circulator prevents leftward-propagating modes from interacting with the \rqu{}, so we ignore them. The internal dissipation in the low-frequency resonator is modeled as arising from a semi-infinite transmission line of characteristic impedance $R_b$. The incident and reflected modes on this transmission line are $\hat{b}_{\rm{in}}$ and $\hat{b}_{\rm{out}}$, respectively.  

The noise fluctuations in these input and output modes can be analyzed using standard input-output theory \cite{Aspelmeyer2014,Walls2008}, treating $\hat{H}_\kappa$ and $\hat{H}_\gamma$ as the microwave and low-frequency bath Hamiltonians, respectively. After making the  Markovian approximation neglecting memory effects in the bath modes, the Heisenburg-Langevin equations of motion for the system are:
\begin{align}
    \label{eq:lab-frame-HL-a}
    \dot{\hat{a}} = \frac{i}{\hbar} \left[\hat{H}_0 + \hat{H}_{\rm{int}}^{\rm{RQU}}, \hat{a}\right] - \frac{\kappa}{2} \hat{a} - \sqrt{\kappa} \hat{a}_{\rm{in}}, \\
      \label{eq:lab-frame-HL-b}
    \dot{\hat{b}} = \frac{i}{\hbar} \left[\hat{H}_0 + \hat{H}_{\rm{int}}^{\rm{RQU}}, \hat{b}\right] - \frac{\gamma}{2} \hat{b} - \sqrt{\gamma} \hat{b}_{\rm{in}}, 
\end{align}
where $\kappa$ is the decay rate of the circulating power in the microwave resonator, and $\gamma$ is the decay rate in the low-frequency resonator. Dots indicate time derivatives. 
$\hat{H}_{\rm{int}}^{\rm{RQU}}$ describes three-wave mixing, so equations \ref{eq:lab-frame-HL-a} and \ref{eq:lab-frame-HL-b} are nonlinear. Our present analysis will focus on the regime where we can linearize $\hat{H}_{\mathrm{int}}$, although experiments in the highly nonlinear regime (when the single-photon interaction rate $g_0$ exceeds the microwave loss rate $\kappa$) are also interesting. To linearize, we write the microwave ladder operators as a sum of a classical amplitude and small quantum fluctuations:
\begin{gather}
    \label{eq:a-classical-plus-quantum}
    \hat{a} = \bar{a} + \delta \hat{a}, \\
    \label{eq:ain-classical-plus-quantum}
\hat{a}_{\rm{in}} = \bar{a}_{\rm{in}} + \delta \hat{a}_{\rm{in}}, \\
    \label{eq:aout-classical-plus-quantum}
\hat{a}_{\rm{out}} = \bar{a}_{\rm{out}} + \delta \hat{a}_{\rm{out}}. 
\end{gather}
Here, $|\bar{a}|^2$ represents the average photon number circulating in the microwave resonator due to the drive. The drive has an amplitude $|\bar{a}_{\rm{in}}|^2$ (in units of photons/second). Likewise, the average photon flux propagating towards the microwave amplifier is given by $|\bar{a}_{\rm{out}}|^2$. The boundary condition relating the output, input, and intra-cavity fields is:
\begin{equation}
    \label{eq:input-output-boundary-condition}
    (\hat{a}_{\rm{out}}-\hat{a}_{\rm{in}}) =  \sqrt{\kappa}\hat{a} 
\end{equation}

We operate the upconverter in the regime of strong microwave drives, such that $|\bar{a}|\gg1$. Inserting expression \ref{eq:a-classical-plus-quantum} into the interaction Hamiltonian \ref{eq:H-RQU-int} yields:
\begin{equation}
    \label{eq:H-int-classical-plus-quantum-expansion}
    \hat{H}_{\mathrm{int}}^{\rm{RQU}} = -\hbar g_0 \left(|\bar{a}|^2 +  \bar{a}^*\delta\hat{a} + \bar{a} \delta \hat{a}^\dagger + \delta\hat{a}^\dagger \delta\hat{a} \right) (\hat{b}^\dagger + \hat{b}).
\end{equation}
The first term represents a constant flux offset applied to the low-frequency resonator, and can be ignored. The second and third terms represent mixing between the coherent microwave amplitudes $\bar{a}^*$ and $\bar{a}$ and the quantum noise terms $\delta \hat{a}$ and $\delta \hat{a}^\dagger$. The last term represents the interaction of the quantum noise with itself, and since it is smaller than the second and third terms by a factor of $1/|\bar{a}| \ll1$, we ignore it. Thus, the relevant portions of the interaction Hamiltonian are: 
\begin{equation}
    \label{eq:eq:H-int-classical-plus-quantum-expansion-APPROX}
    \hat{H}_{\rm{int}}^{\rm{RQU}} \approx -\hbar g_0 \left( \bar{a}^*\delta\hat{a} + \bar{a} \delta \hat{a}^\dagger    \right) (\hat{b}^\dagger + \hat{b}).
\end{equation}
Since $\bar{a}$ is a classical value rather than an operator, this interaction now describes an effective two-wave interaction with a tunable strength set by $g_0 |\bar{a}|$, rather than the three-wave interaction described by equation \ref{eq:H-RQU}. This approximation holds for strong microwave drivesHence, we refer to it as the linearized interaction.

In order for $\bar{a}$ to have a non-zero value in the steady state, external energy must be supplied to the microwave resonator via a driving term. A monotonic, coherent drive tone applied via the readout waveguide induces the drive Hamiltonian:
\begin{equation}
    \label{eq:H-drive}
       \hat{H}_{\rm{drive}} = -i \hbar \sqrt{\kappa} (\Bar{a}_{\rm{in}}(t)\hat{a}^\dagger + \Bar{a}_{\rm{in}} ^*(t) \hat{a}),
\end{equation}
where $\Bar{a}_{\rm{in}}(t)$ represents the (classical) amplitude of the coherent voltage drive applied to the microwave resonator via the input transmission line. We apply the unitary transformation:
\begin{equation}
    \label{eq:rotating-frame-unitary}
    \hat{H}_{\rm{rot}} = \hat{U}\hat{H}_{\rm{lab}}\hat{U}^\dagger - i \hat{U}\dot{\hat{U}}^\dagger,
\end{equation}
to move to a frame rotating at the drive frequency. $\hat{U}=e^{(i \omega_d \hat{a}^\dagger \hat{a}t)}$, where $\omega_d$ is the angular velocity of the rotating frame. We use $\hat{H}_{\mathrm{lab}} =  \hat{H}_0+\hat{H}_{\mathrm{int}}^{\mathrm{RQU}}$ as the lab-frame Hamiltonian. In this frame, the Heisenberg-Langevin equation of motion for $\hat{a}$ reads:
\begin{equation}
    \label{eq:rotating-frame-HL-a}
    \dot{\hat{a}} = \frac{i}{\hbar} \left[-\hbar\left( \Delta  + g_0(\hat{b}^\dagger + \hat{b})\right) \hat{a}^\dagger \hat{a}  , \hat{a}\right] - \frac{\kappa}{2} \hat{a} - \sqrt{\kappa} \hat{a}_{\rm{in}},
\end{equation}
where $\Delta \equiv \omega_{d}-\omega_a$.

We begin by solving for the steady-state amplitude $\bar{a}$ using the classical portion of the equation of motion \ref{eq:rotating-frame-HL-a}, neglecting terms of order $\delta \hat{a}$ and with no flux applied from the sensor: $\hat{\Phi}=0$. We find:
\begin{equation}
    \label{eq:phase-insensitive-classical-amplitude}
    \bar{a} = \frac{\sqrt{\kappa}\bar{a}_{\rm{in}}}{ i\Delta -\frac{\kappa}{2}}.
\end{equation}

We can insert this steady state classical amplitude into equations \ref{eq:lab-frame-HL-b} and \ref{eq:rotating-frame-HL-a} in order to describe the linearized dynamics of the coupled modes in the frequency domain.  After neglecting DC terms and small terms of order $\delta \hat{a} \hat{b}$, we find:
\begin{multline}
    \label{eq:da-EOM}
    -i\omega \delta\hat{a}[\omega] =  (i\Delta - \kappa/2 ) \delta \hat{a}[\omega] - \\ i g_0 \bar{a} (\hat{b}^\dagger[\omega] + \hat{b}[\omega])  - \sqrt{\kappa} \delta \hat{a}_{\rm{in}}[\omega] \\
\end{multline}
for the intra-cavity field, and
\begin{multline}
    \label{eq:b-EOM}
    -i\omega \hat{b}[\omega]  = (-i \omega_b - \gamma/2)\hat{b }[\omega] - \\i g_0 (\bar{a}^* \delta \hat{a}[\omega] + \bar{a} \delta \hat{a}^\dagger[\omega])  - \sqrt{\gamma} \hat{b}_{\rm{in}}[\omega] 
\end{multline}
for the low-frequency mode.

Equations \ref{eq:da-EOM} and \ref{eq:b-EOM} fully describe the dynamics of the \rqu{} in the linearized regime, and can be used to calculate the behavior of the coupled modes in a variety of regimes. The approximations used in this analysis hold in the regime of strong microwave drives ($|\bar{a}|\gg 1$) and moderate interaction strength ($g_0<\kappa$), and they simplify the dynamics of the coupled system of resonators to an effective two-wave interaction with tunable strength. As discussed in the next section, this description allows the \rqu{} system to be analyzed as a linear, op-amp mode amplifier. The quantum metrology technique in \S\ref{sec:TwoToneDrives} arises from a special cases of these linearized dynamics where microwave drive consists of a superposition of drive tones at $\Delta = \pm \omega_b$.

\subsection{Quantum amplifier theory}
\label{subsec:op-amp}

When the \rqu{} is driven by a single microwave tone, it functions as a phase-preserving \cite{caves1982quantum} electromagnetic amplifier that measures both the low-frequency signal quadratures with equal sensitivity. It is an op-amp mode amplifier \cite{Clerk2010} in the sense that it measures an input state variable (flux) in a lumped-element circuit rather than a traveling wave on  a transmission line. The \rqu{} maps the flux variable $\hat{\Phi}$ of the low-frequency circuit onto the microwave output mode $\delta \hat{a}_{\rm{out}}$. 

The upconversion process adds noise as required by the Standard Quantum Limit on amplification, and in this section we show that the \rqu{} can achieve readout at the SQL \cite{Clerk2010}. In this readout protocol, the microwave readout tone is resonant ($\Delta=0$), and the low frequency signal appears in $\delta \hat{a}_{\rm{out}}$ as symmetric sidebands due to the phase modulation of the reflected microwave signal, which is a three-wave mixing process.

In order to evaluate the total noise added in the upconversion process, we calculate fluctuations in the output mode $\delta \hat{a}_{\rm{out}}$. We use the boundary condition in equation \ref{eq:input-output-boundary-condition} to eliminate the intra-cavity field in equation \ref{eq:da-EOM}, yielding the  equation of motion governing the small quantum fluctuations of the input and output microwave modes, and the low frequency mode:
\begin{equation}
    \label{eq:daout-eom}
    \delta \hat{a}_{\rm{out}}[\omega] = \frac{i \omega -\kappa/2}{i \omega 
    +\kappa/2} \delta \hat{a}_{\rm{in}}[\omega] + \frac{i g_0 \bar{a}\sqrt{\kappa}}{(i \omega   + \kappa/2)\Phi_{\rm{ZPF}}} \hat{\Phi}[\omega].
\end{equation}
The first term on the RHS of equation \ref{eq:daout-eom} represents the fluctuations in the input mode, which are reflected from the microwave resonator with a phase shift, but no change in amplitude. These fluctuations carry no information about the state of the low-frequency resonator, and cause uncertainty in  $\hat{\Phi}$, referred to as imprecision noise. The second term carries information about the state of the low-frequency resonator encoded in sidebands at $\pm \omega$ (in the rotating frame).   

The other irreducible noise source arises from fluctuations in the intra-cavity field $\delta \hat{a}$ which perturb the state of the low-frequency resonator. Inserting the steady state solution for the intra-cavity field into equation \ref{eq:lab-frame-HL-b} and focusing on the backaction terms yields an equation of motion for the low-frequency mode:
\begin{multline}
    \label{eq:b-eom}
    \hat{b}[\omega] = 
    \left(i\omega_b-i\omega+\frac{\gamma}{2}\right)^{-1}  \frac{i g_0\bar{a}}{\sqrt{\kappa}} \left(\delta\hat{a}^\dagger[\omega] + \delta\hat{a}[\omega]\right),
\end{multline}
where (without loss of generality) we have set the phase of $a_{\rm{in}}$ so that $\bar{a}$ is real. Equation \ref{eq:b-eom} captures the effects of microwave fluctuations that perturb the state of the low-frequency resonator, including backaction due to fluctuations in the microwave field (proportional to $\delta \hat{a}$ and $\delta \hat{a}^\dagger$). Together, equations \ref{eq:daout-eom} and \ref{eq:b-eom} describe the two irreducible noise sources in the upconversion process. Since the \rqu{} is functioning as an op-amp (meaning that it measures a state variable rather than a traveling wave), it is more convenient to describe the noise in the upconversion process directly in terms of the voltages and currents in the low-frequency resonator.

We take the limit of low frequencies, where the generated sidebands are within the bandwidth of the microwave resonator, $\omega \ll \kappa/2$. The total output signal at the follow-on microwave amplifier is given by equation \ref{eq:aout-classical-plus-quantum}. In order to recover the flux signal, we noiselessly amplify (with a degenerate JPA) and demodulate using a reference tone at $\omega_d$, measuring the microwave phase quadrature $\hat{a}_{\rm{out}}[\omega] - \hat{a}_{\rm{out}}^\dagger[\omega]$:
\begin{equation}
    \label{eq:detection-quadrature-at-JPA}
    \begin{split}
    \hat{a}_{\rm{out}}[\omega] - \hat{a}_{\rm{out}}^\dagger[\omega] =  \delta \hat{a}_{\rm{in}}[\omega] - \delta \hat{a}_{\rm{in}}^\dagger[\omega] 
    + \frac{4i g_0 \bar{a}}{\sqrt{\kappa}} \frac{\hat{\Phi}[\omega]}{\Phi_{\rm{ZPF}}}.
    \end{split}
\end{equation}

The fluctuating terms ($\delta \hat{a}_{\rm{in}}[\omega] - \delta \hat{a}_{\rm{in}}^\dagger[\omega]$) in the microwave phase quadrature cause imperfect reconstruction of the low-frequency current at the input of the \rqu{} after microwave amplification and demodulation, which we refer to as imprecision noise. This imprecision noise can be referred back to input currents with equation \ref{eq:detection-quadrature-at-JPA}, yielding:
\begin{equation}
    \label{eq:Phi-imp}
    \hat{I}_{imp} =  \frac{\Phi_{\rm{ZPF}}\sqrt{\kappa}}{4 i g_0 M \bar{a}} \left( \delta\hat{a}_{\rm{in}}[\omega]  -\delta\hat{a}_{\rm{in}}^\dagger[\omega] \right),
\end{equation}
where $M$ is the effective mutual inductance relating the input flux signal $\hat{\Phi}$ to $\hat{I}$, the current flowing through $L_b$.

Using the backaction terms in equation \ref{eq:b-eom}, we can write the perturbation of the low-frequency current due to backaction. We find:
\begin{equation}
    \label{eq:I-from-Ohms-Law}
    \hat{I}_{BA}[\omega] = \left(Y_+[\omega] + Y_-[\omega]\right) \hat{V}_{BA}[\omega],
\end{equation}
where  $Y_\pm[\omega]$ is approximately the admittance of the low-frequency resonator at its positive and negative resonance frequencies:
\begin{equation}
    \label{eq:Ypm}
    Y_\pm[\omega] = \frac{1}{2iL_b(i\gamma/2 -\omega \pm \omega_b )}
\end{equation}
The approximation in equation \ref{eq:Ypm} holds for frequencies near the resonance frequencies $\omega\approx\pm\omega_b$. In order to evaluate the effect of backaction at frequencies very detuned from the resonance frequency, the rotating wave approximation implicit in the derivation of equation \ref{eq:lab-frame-HL-b} would have to be dropped. The backaction voltage $\hat{V}_{BA}[\omega]$ arises from the fluctuation terms $\delta \hat{a}$ and $\delta \hat{a}^\dagger$, and is given by:
\begin{equation}
    \label{eq:Vba}
    \hat{V}_{BA}[\omega] = \frac{8 i\omega M g_0 Q_{\rm{ZPF}}  |\bar{a}| }{\kappa}( \delta\hat{a}[\omega] +  \delta\hat{a}^\dagger[\omega]),
\end{equation}

Equations \ref{eq:Vba} and \ref{eq:Phi-imp} show the  backaction and imprecision noise contributions arise from the input quadratures $\hat{\delta}_{BA} = \delta\hat{a}_{\rm{in}}[\omega] +\delta\hat{a}_{\rm{in}}^\dagger[\omega]$ and  $\hat{\delta}_{imp} = \delta\hat{a}_{\rm{in}} [\omega] -\delta\hat{a}_{\rm{in}}^\dagger[\omega]$, respectively. 
If the input is prepared in a coherent state without additional noise and the output mode is detected without adding noise (e.g. if $\delta\hat{a}_{\rm{in}}, \delta\hat{a}_{\rm{in}}^\dagger$, are sourced from a cold resistor, and a noiseless, phase-sensitive JPA detects  $\delta\hat{a}_{\rm{out}}, \delta\hat{a}_{\rm{out}}^\dagger$), the fluctuations of these mode quadratures are described by the (symmetrized) noise spectral densities:
\begin{gather}
    \label{eq:S_deltaBA}
    \bar{S}_{\delta_{BA}\delta_{BA}} [\omega]= 1, \\
    \label{eq:S_deltaimp}
    \bar{S}_{\delta_{imp}\delta_{imp}}[\omega] = 1, \\
    \label{eq:S_deltaBAdeltaimp}
    \bar{S}_{\delta_{BA}\delta_{imp}}[\omega] = 0, 
\end{gather}
In other words, the quadratures have uncorrelated fluctuations with a total amplitude corresponding to a single quantum. We can now evaluate the imprecision and backaction spectral densities:
\begin{gather}
    \label{eq:S_II}
    \bar{S}_{II} = \frac{\Phi_{\rm{ZPF}}^2 \kappa}{8 |\bar{a}|^2 g_0^2 M^2}, \\
     \label{eq:S_VV}
    \bar{S}_{VV} = \frac{8|\bar{a}|^2 g_0^2M^2\omega^2 Q_{\rm{ZPF}}^2}{\kappa}, \\
    \bar{S}_{IV} = 0.
\end{gather}
These spectral densities describe an op-amp mode amplifier capable of operating at the Standard Quantum Limit for a phase-preserving amplifier. The noise impedance of this amplifier is tunable without changing the geometry or input inductance of the device, simply by changing the microwave amplitude $|\bar{a}|$. The total noise added by such an amplifier is the sum of these three contributions, with the op-amp achieving a total added current noise of:
\begin{equation}
     \bar{S}_{II,tot} = \bar{S}_{II} + \bar{S}_{VV} |\bar{Y}[\omega]|^2 + 2 \mathrm{Re}\left(\bar{S}_{IV}\bar{Y}[\omega]^*\right),
\end{equation}

\noindent
where $\bar{Y}[\omega]=Y_+[\omega] + Y_-[\omega]$ is the sum of the positive- and negative-frequency components of the admittance  On resonance at $\omega=\omega_b$, the admittance is purely real: $\bar{Y}[\omega_b]=1/R_b$. Using also the fact that $S_{IV} =0$, we can simplify the total noise to:
\begin{equation}
    2 k_B T_N[\omega_b] = \frac{\bar{S}_{VV}}{R_b} + R_b \bar{S}_{II},
\end{equation}\label{eq:TotalNoise}
where $T_N[\omega]$ is the noise temperature, which is a function of the frequency, evaluated on resonance at $\omega=\omega_b$. Note that the densities in equations \ref{eq:S_II} and \ref{eq:S_VV} have different scalings with respect to the microwave drive power $|a|$. Thus, for the simplified case of optimizing the added noise on resonance, with $\Delta=0$, we find that the real, on-resonance input circuit resistance $R_{\rm noise}$ (the ``noise impedance'') that optimizes the noise temperature is
\begin{equation}
    R_{\rm noise}=\sqrt{\frac{S_{VV}}{S_{II}}}
    =\frac{8 |\bar{a}|^2 g_0^2 \omega   M^2}{\kappa} \frac{ Q_{\rm{ZPF}}}{ \Phi_{\rm{ZPF}}}.
    \label{eq:NoiseResistance}
\end{equation} 
\noindent
So, the noise resistance of the RQU can be tuned by changing the pump power  $|\bar{a}|^2$, without changing the input inductance.

The noise temperature is optimized when $R_b=R_{\rm noise}$. We find the optimal power level is given by:
\begin{equation}
    \label{eq:abar-optimal}
        |\bar{a}|^2 = \frac{R_b  \kappa}{ 8 g_0^2 M^2 \omega }\frac{ Q_{\rm{ZPF}}}{ \Phi_{\rm{ZPF}}},
\end{equation}
with an overall noise temperature of:
\begin{equation}
    \label{eq:noise-temp-optimal}
    k_B T_{N}[\omega_b] = \frac{\hbar \omega_b}{2}
\end{equation}
This corresponds to an op-amp mode amplifier operating at the SQL. Electromagnetic measurement techniques operating at the SQL are available at higher microwave frequencies, but are not presently available for signal frequencies at lower frequencies between dc and VHF. The \rqu{} thus adds valuable capability to the toolbox of techniques available for quantum metrology. 

In reality, we cannot operate the \rqu{} at arbitrarily high microwave drive levels. The upper limit on the value of $|\bar{a}_{\rm{in}}|^2$ will  be set by nonlinear terms in the Josephson inductance, leading to Kerr-type nonlinearity, where the effective inductance of the interferometer is a function of amplitude of the currents circulating in the microwave resonator. This introduces terms of the form:
\begin{equation}
    \hat{H}_{\rm Kerr} = \hbar \Lambda \hat{a}^\dagger \hat{a}^\dagger \hat{a}\hat{a},
\end{equation}
where $\Lambda$ quantifies the strength of the quartic nonlinearity in the microwave circuit. Kerr-type nonlinearity has been extensively studied in the context of parametric amplifiers, but the most relevant effect for the \rqu{} is the microwave power-dependent frequency shift. At high microwave powers, this frequency shift causes the microwave resonator to bifurcate. This imposes an upper limit on the microwave power circulating within the \rqu{}:
\begin{equation}
    |\bar{a}_{\rm{max}}|^2 = \chi |\bar{a}_{bif}|^2 =\frac{1}{2\sqrt{3}}\frac{\kappa}{\Lambda},
\end{equation}
where $\chi$ is a dimensionless parameter chosen by the operator, which quantifies how far below the onset of bifurcation the \rqu{} is operated, with $\chi \ll 1$ defining the regime of linear microwave response. $|\bar{a}_{bif}|^2$ is the average circulating photon number at the onset of bifurcation \cite{LaFlamme2010QauntumLimitedAmplification}. This  limits the highest noise resistance $R_{\rm{max}}$ that can be achieved: 
\begin{equation}
    R_{\rm max}=\sqrt{\frac{S_{VV}}{S_{II}}}
    =\frac{8 |\bar{a}_{\rm{max}}|^2 g_0^2 \omega   M^2}{\kappa} \frac{ Q_{\rm{ZPF}}}{ \Phi_{\rm{ZPF}}}. 
    \label{eq:RMax}
\end{equation}

\section{Measuring untuned input circuits with upconversion}
\label{sec:untuned}

In some applications, especially at low frequencies, resonant input circuits are not practical. Instead, a flux signal is measured in an untuned inductive load such as a magnetometer coil. The sensitivity of this readout is often quantified by its imprecision, uncoupled ``energy sensitivity,'' which expresses the smallest current signal that can be detected above the imprecision noise for a given inductance \cite{Voss1981}. In the case of an untuned circuit, the impedance of the input circuit is so high that backaction noise is insignificant. In this case, the only limit on the noise is set by the fluctuations in the imprecision quadrature $\hat{\delta}_{imp}$, and the final noise temperature is determined by the precision of the phase estimation, as determined by the standard quantum limit on interferometry \cite{PhysRevLett.99.223602}.

We can express the energy sensitivity of an upconverter operated with an untuned input circuit as:
\begin{equation}
\label{eq:epsilon}
    \epsilon = \frac{\bar{S}_{II} L_b}{2},
\end{equation} 
where $L_b$ is the self inductance of the input coil which couples flux from the low-frequency circuit into the interferometer. The imprecision energy sensitivity can be expressed as a multiple of $\hbar$. The details of the operation of dc SQUIDs limits their imprecision energy sensitivity to $\epsilon \gtrsim \hbar$ (see for example \cite{koch1981}), but we emphasize that this is not a standard quantum limit. It is nonetheless the appropriate figure of merit for important applications with untuned input circuits.

In a dc SQUID, the input imprecision current noise can be reduced by increasing the inductance, but this does not change the imprecision energy sensitivity \ref{eq:epsilon}. However, in an RQU, the imprecision current noise can be reduced by increasing the pump power $|\bar{a}|^2 $, without changing the input inductance, reducing the imprecision energy sensitivity.

Substituting equation \ref{eq:S_II} into equation \ref{eq:epsilon}, we see that at low frequencies ($\omega \ll \kappa/2$), the \rqu{} achieves an imprecision-noise-limited uncoupled energy sensitivity $\epsilon < \hbar$ as long as the amplitude of the microwave drive is larger than:
\begin{equation}
    |\bar{a}|^2 > \frac{\omega_b \kappa L_b^2}{16 g_0^2 M^2},
\end{equation}

\noindent
suggesting that it may outperform the best dc SQUIDs in some untuned applications. To achieve this performance, the RQU must be designed so that this tone power can be applied without approaching junction critical currents or resonator bifurcation \cite{Manucharyan2007}.

\section{Numerical Estimates of RQU Parameters}
\label{sec:NumericalEstimates}

In order to inform the design of a practical \rqu{}, we introduce an example low frequency resonator. This example resonator has parameters similar to those that are useful in a variety of precision measurement tasks, e.g. searches for low-mass dark matter candidates \cite{chaudhuri2018} or high sensitivity measurements of nuclear spin ensembles \cite{clarkeSpinNoise}. There is substantial flexibility in the design of the RQU which would allow the design to be adapted to higher or lower frequency ranges, or broadband experiments, but the example resonator serves as a benchmark for initial RQU designs. 

In order to reach the SQL near the low-frequency resonance, the RQU must be driven with a sufficiently strong microwave tone such that its noise impedance (given by equation \ref{eq:NoiseResistance}, with a maximum value set in \ref{eq:RMax}) can match the source impedance of the low frequency resonator, which is entirely determined by its inductance, resonant frequency, and quality factor. We choose a 5\;\textmu H inductance as a representative value for a centimeter-scale pickup coil with a few tens of turns (as might be used in a low-mass dark matter experiment searching for electromagnetic \cite{phipps2020} or nuclear \cite{aybas} interactions). We also note that in principle it is possible to use an $N$-turn step down transformer to lower the pickup coil impedance seen by the RQU by a factor of $1/N^2$. We do not include this transformer here when evaluating RQU designs, but a transformer might be used to increase the noise impedance further than is achievable with a practical microwave drive tone.

Precisely on-resonance, the impedance of the low-frequency resonator is real, with value
\begin{equation}
    R_b = \frac{\omega_b L_b}{Q_b}.
\end{equation}
In order to reach the SQL, the maximum noise impedance of the RQU must exceed $R_b$, so that the optimal microwave drive level given in equation 42 can be reached without bifurcation:
\begin{equation}
    R_{\rm max} > \frac{\omega_b L_b}{ Q_b}.
\end{equation}
We can also express this condition as a minimum quality factor $Q_{\rm min}$ which the low-frequency resonator must achieve in order to be read out at the SQL:
\begin{equation}
    Q_{\rm min} = \frac{\omega_b L_b}{R_{\rm max}}. 
\end{equation}
For a fiducial RQU design based on a SQUID with an effective critical current $I_c =5$\;\textmu A, we estimate $R_{\rm max}$ to  be 500 \textmu$\Omega$, resulting in $Q_{\rm min}\approx 62,000$, a reasonable Q for a superconducting resonator constructed with low-loss materials in this frequency range, e.g. \cite{KuenstnerThesis}.


\section{Two-Tone Microwave Drives}
\label{sec:TwoToneDrives}

The single-tone microwave drive scheme above makes the \rqu{} operate as a linear, phase-insensitive op-amp, subject to the SQL on amplification. Quantum metrology, including measurements better than the SQL, are enabled by more sophisticated drive schemes. In this section we analyze the \rqu{} when the microwave drive signal consists of two tones, symmetrically detuned above and below the microwave resonance, which can be used to implement quantum backaction evasion.  

The two-tone microwave drive is given by:
\begin{equation}
    \bar{a}_{\rm in} = a_{\rm drive} \sin(\omega_b t + \phi_{\rm drive}) e^{i \omega_a t}, 
\end{equation}
where $a_{\rm drive}$ specifies the amplitude of the two-tone drive, and $\phi_d$ sets the phase of the amplitude modulation. Without loss of generality, we can set $\phi_{\rm drive}=0$.  
Solving for the classical amplitude of the field within the microwave resonator yields: 
\begin{equation}
    \bar{a} = a_{\rm mod} \cos (\omega_b t + \delta) e^{-i \omega_a t},
\end{equation}
where $a_{\rm{mod}}$ captures the peak amplitude of the modulated microwave tone:
\begin{equation}
    \label{eq:a_max}
    a_{\rm mod} = a_{\rm drive} \sqrt{\frac{\kappa}{\kappa^2 + 4 \omega_b^2}},
\end{equation}
and $\delta$ encodes the phase of the amplitude modulation envelope:
\begin{equation}
    \label{eq:delta}
    \delta = \arctan (\kappa/\omega_b).
\end{equation}

Since the \rqu{} biased this way functions as a phase-sensitive amplifier, it is more useful to define the state of the low-frequency resonator by its quadrature operators:
\begin{align}
    \label{eq:x_quadrature}
    \hat{X} = \frac{1}{\sqrt{2}} (\hat{b} e^{i \omega_b t} +  \hat{b}^\dagger e^{-i \omega_b t}  ),\\
    \label{eq:y_quadrature}
    \hat{Y} = \frac{-i}{\sqrt{2}} (\hat{b} e^{i \omega_b t} +  \hat{b}^\dagger e^{-i \omega_b t}  )
\end{align}

We can calculate the equations of motion for $\hat{X}$ and $\hat{Y}$, following the derivation in \cite{Clerk2008}. We find that the $\hat{Y}$ quadrature suffers measurement backaction, with noise spectral density given by: 
\begin{align}
    \label{eq:S_YY}
    \bar{S}_{YY} [\omega] =  \frac{\gamma/2}{\omega^2 + (\gamma/2)^2} \left( 1 + 2(n_{\rm eq} + n_{\rm BA} + n_{\rm bad}) \right),
\end{align}
where $n_{\rm eq}$ describes the thermal occupation of the low-frequency resonator 
and $n_{\mathrm{BA}}$ is the microwave backaction on the strongly-perturbed $\hat{Y}$ quadrature, and $n_{\rm bad}$ is a spurious backaction term which acts on both quadratures equally. In contrast, the $\hat{X}$ quadrature suffers reduced backaction, with a total noise spectral density (after accounting for imprecision noise added by the microwave readout) of: 
\begin{align}
    \label{eq:S_XX}
    \bar{S}_{XX} [\omega] =  \frac{\gamma/2}{\omega^2 + (\gamma/2)^2} \left( 1 + 2(n_{\rm eq} + n_{\rm bad} + n_{\mathrm{imp}}) \right),
\end{align}
where $n_{\mathrm{imp}}$ represents imprecision noise in the measurement due to noise in the microwave drive and readout. Note that the microwave signal carries no information about the $\hat{Y}$ quadrature, so equation \ref{eq:S_YY} represents only the \emph{internal} fluctuations of the low-frequency resonator, while equation \ref{eq:S_XX} includes both the internal fluctuations and the additional uncertainty in the $\hat{X}$ quadrature after microwave detection. 

In terms of photon numbers, the various noise terms are given by:
\begin{align}
    n_{\rm{eq}} = \left(e^{\hbar \omega_b/k_B T
    }-1\right)^{-1}, \\ 
    n_{\rm{BA}} =  \frac{2(a_{\mathrm{mod}}g_0 \Phi_{\mathrm{ZPF}})^2 }{\kappa\gamma},\\
    n_{\rm{bad}} = \frac{(a_{\rm{mod}} g_0 \Phi_{\rm{ZPF}})^2}{16 \kappa \gamma} \left(\frac{\kappa}{\omega_b}\right)^2, \\
    n_{\rm{imp}} = \frac{\kappa 
    \gamma S_0}{32 (a_{\mathrm{mod}} g_0 \Phi_{\mathrm{ZPF}} )^2},
\end{align}
where $T$ is the equilibrium temperature of the low-frequency resonator, and $S_0$ is the noise associated with the microwave readout, with $S_0=1$ corresponding to shot-noise limited microwave readout (via, for example, a phase-insensitive JPA operating at the SQL).

The total added noise on the $\hat{X}$ quadrature (not including the equilibrium thermal fluctuations $n_{\mathrm{eq}}$) is:
\begin{align}
    n_{\mathrm{add, X}} =  n_{\rm{bad}} + n_{\rm{imp}},
\end{align}
with the SQL corresponding to $n_{\mathrm{add, X}} = 1/2$. The strength of the measurement can be parametrized by $n_{\rm{BA}}$, with the quadrature noises equivalent to:
\begin{align}
    n_{\mathrm{add, X}} =  \frac{n_{\rm{BA}}}{32}\left(\frac{\kappa}{\omega_b}\right)^2 + \frac{S_0}{16 n_{\rm{BA}} },
\end{align}

In the limit of good sideband resolution ($\omega_b/\kappa \gg 1$), this spurious backaction can be arbitrarily suppressed, corresponding to a backaction-free, or quantum non-demolition (QND) measurement. For shot-noise limited readout ($S_0=1$), the total noise in the $\hat{X}$ quadrature will evade the SQL as long as the measurement strength reaches $n_{\rm{BA}}\geq 1/8$. Importantly, this backaction evasion scheme can still perform better than the SQL with finite sideband resolution and with non-ideal microwave readout noise. For example, for a very modest sideband resolution  ($\omega_b/\kappa=2$) and microwave readout noise of a typical HEMT amplifier ($S_0\approx32$), $n_{\mathrm{add, X}}<1/2$ as long as the measurement strength exceeds $n_{\rm{BA}}\approx 4.3$. 

In this scheme, the \rqu{} can operate at high microwave power levels, reducing imprecision noise on the $\hat{X}$ quadrature, without the backaction penalty present in the single-tone, phase-insensitive mode of operation (as captured by the backaction voltage term in equation \ref{eq:S_VV}). This backaction-free measurement mode could allow a quantum speedup in dark matter searches.

\section{\rqu{} Upconversion Demonstration}
\label{sec:upconversion-demo}

\begin{figure}
\includegraphics[width=0.95\linewidth]{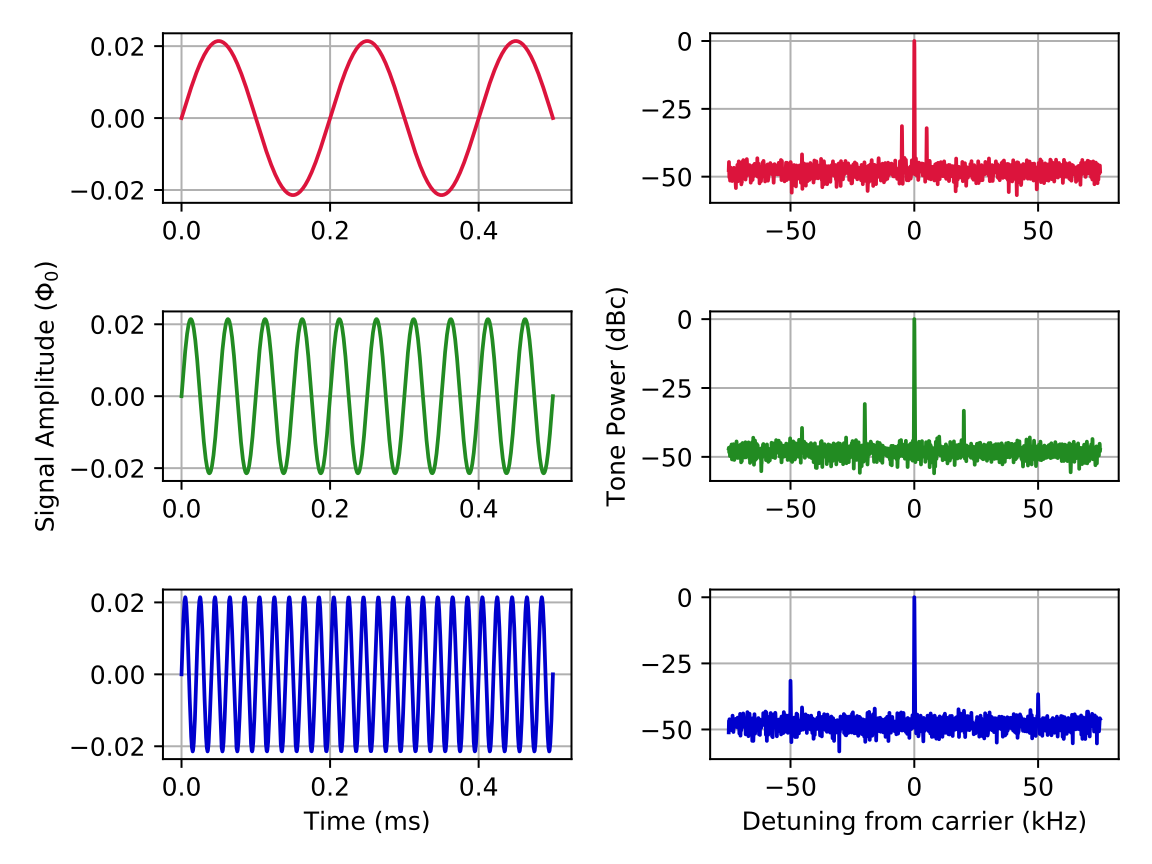}
\caption{Data demonstrating phase-insensitive upconversion, showing the strong carrier tone used to bias the \rqu{} at $\Delta\sim0$ and the two weak signal sidebands, for a variety of different frequencies of the flux signal $\Phi_b$.}
\label{fig:phase-insensitive}
\end{figure}

\begin{figure}
\includegraphics[width=0.95\linewidth]{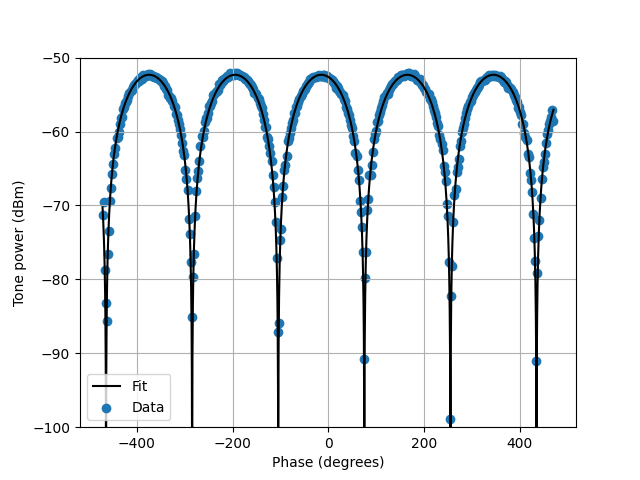}
\caption{A demonstration of phase-sensitive upconversion, with a 46.9\,dB phase-sensitive extinction ratio. The data are fit to a model with only the amplitude and phase as free parameters.}
\label{fig:phase-sensitive-gain-data}
\end{figure}

We demonstrate RQU operation as both a phase-insensitive amplifier (with one tone on resonance, shown schematically in figure \ref{fig:phase-insensitive}), and as a phase-sensitive amplifier (with two pump tones symmetrically detuned on either side of the microwave resonance, shown in figure \ref{fig:phase-sensitive-gain-data}), using a quarter-wave microwave resonator with a single-junction SQUID at the current antinode providing flux tunability.

In order to demonstrate phase-sensitive readout, two tones are synthesized in separate microwave generators and combined in a power splitter. A shared 10MHz clock source allowed the microwave synthesizers, a low-frequency function generator, and a microwave spectrum analyser to generate and detect phase-coherent tones, which drive the upconverter.

\begin{figure}[ht!]
\includegraphics[width=0.7\linewidth]{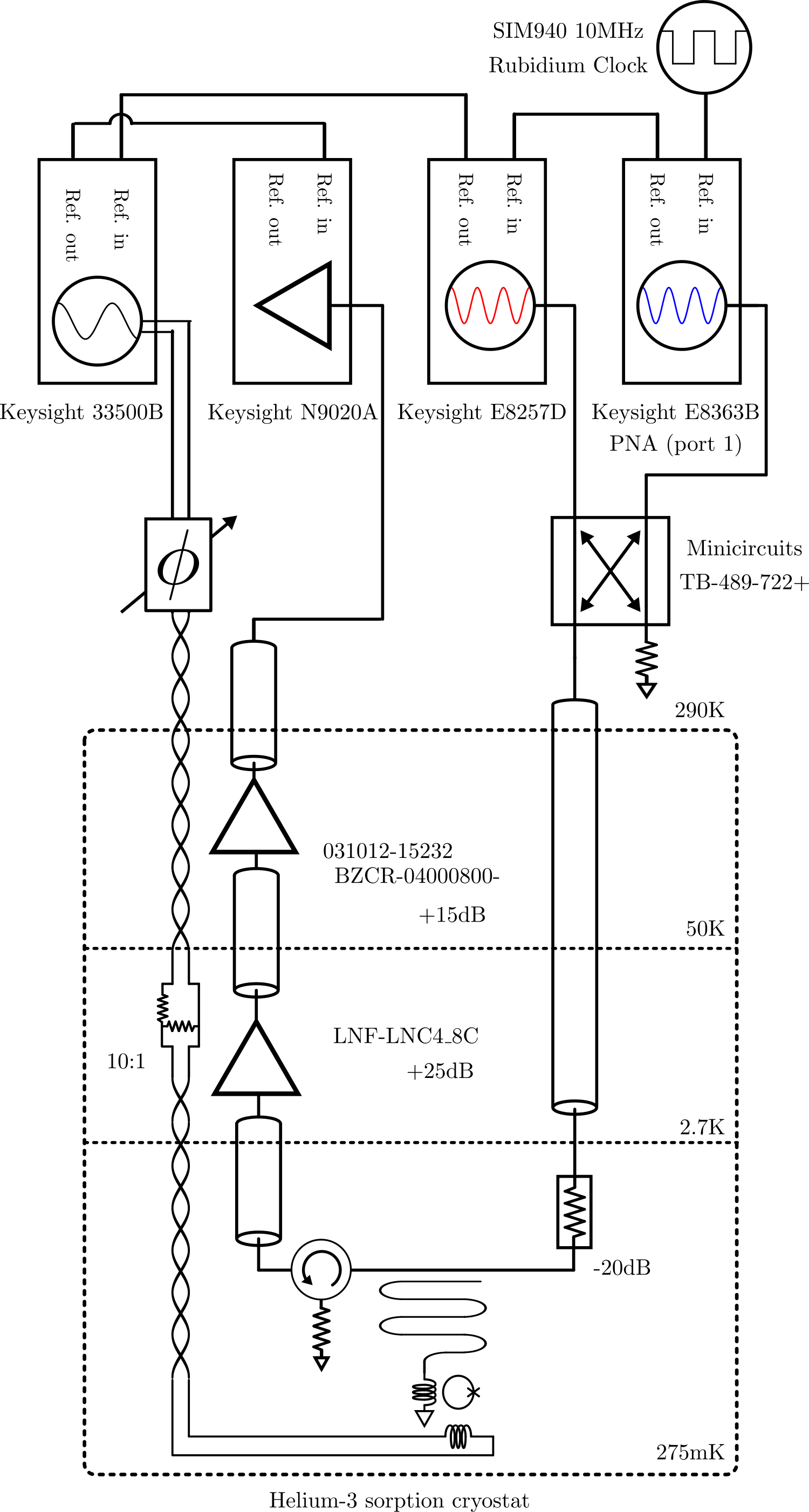}   
\caption{The microwave and cryogenic setup used to demonstrate phase-sensitive upconversion. The \rqu{} is at the base stage of a helium-3 sorption cryostat, with a 4-8GHz microwave readout chain.}
\label{fig:phase-sensitive-setup}
\end{figure}

Figure \ref{fig:phase-sensitive-setup}, shows a schematic of the setup, in which an \rqu{} is operated at T$\sim$300mK at the base stage of a $^3$He sorption cryostat. Filtered and attenuated microwave lines allow for low-noise microwave probe tones to interrogate the \rqu{}, and a High Electron Mobility Transitor (HEMT) amplifier provides low-noise amplification of the tones that transmit past the \rqu{}. A filtered and attenuated twisted-pair line provides flux bias to the SQUID loop of the \rqu{}, allowing for signals up to a few MHz. 

We generate tones symmetrically detuned by 2.9MHz from the 4.89 GHz resonance frequency of the upconverter. We also use the flux bias to modulate the SQUID at 2.9 MHz, and sweep the phase of the SQUID modulation tone over approximately 1080
degrees. The spectrum of the transmitted microwave tones is recorded at a spectrum analyzer. As the phase of the flux modulation changes with respect to the envelope defined by the beating of the microwave tones changes, the total power upconverted modulates, showing a phase-sensitive extinction ratio of 46.9 dB. 

The high extinction ratio proves the viability of phase-sensitive upconversion, although it does not constitute a full backaction-evading measurement, which will require a high-Q resonant circuit on the input of the upconverter. For this signal frequency and this upconverter, the spurious backaction terms $n_{\rm{bad}}$ would have limited the degree of backaction evasion to less than 10dB, in any case. However, reducing the microwave loss rate $\kappa$ by increasing the microwave quality factor can reduce the spurious backaction, in principle arbitrarily.

\section{Conclusions}
\label{sec:conclusions}
We have demonstrated that the \rqu{} is a valuable tool for quantum metrology of low-frequency electromagnetic modes. We have shown that this interaction allows the \rqu{} to act as a quantum-limited op-amp with tunable noise impedance. Using the \rqu{} as a replacement for dc SQUIDs enables amplification at the SQL and \emph{in situ} tunable noise impedance. Furthermore, the \rqu{} can achieve an uncoupled ``energy sensitivity'' $(1/2) L_{\rm in} S_{\rm I}$ significantly below $\hbar$ when coupled to an untuned input inductor. Such performance is not possible with a dc SQUID in which $L_{\rm in}$ is linked to $S_{\rm I}$ \cite{Voss1981}. Since it is capable of operating at the SQL, with tunable noise impedance, and achieving sub-$\hbar$ uncoupled energy sensitivity the \rqu{} exceeds the capability of the current state-of-the-art, dc SQUIDs. The capabilities provided by the \rqu{} can be useful in a variety of magnetometry applications.

The \rqu{} can also be operated as a phase-sensitive amplifier, enabling performance beyond the SQL via backaction evading measurements, which has the potential to dramatically enhance the performance of an important class of fundamental physics experiments. We have demonstrated the basic functionality of upconversion in both phase-insensitive and phase-sensitive modes, converting signals from 5kHz to 3MHz into the microwave C band. The phase-sensitive data has an extinction ratio of 46.9dB, which 
is a necessary step towards achieving a high degree of backaction evasion in future experiments. This will enable beyond-SQL metrology in a variety of precision experiments, including searches for sub-\textmu eV axion dark matter.\\


\section{Acknowledgments}
This work was supported by the US Department
of Energy, Office of High Energy Physics program under the QuantISED program, FWP 100667. S. Chaudhuri acknowledges the support of the R.H. Dicke Postdoctoral Fellowship and the David Wilkinson Fund. CY was supported in part by the National Science Foundation Graduate Research Fellowship Program under
Grant No. 1656518. Part of this work was performed at the Stanford Nano Shared Facilities (SNSF)/Stanford Nanofabrication Facility (SNF), supported by the National Science Foundation under award ECCS-2026822. Additional microfabrication support was provided by K. Multani, A.Y. Cleland and the Safavi-Naeini group.

\bibliographystyle{apalike}
\bibliography{ref}

\end{document}